\date{}
\documentclass[12pt,a4paper]{article}
\usepackage[latin1]{inputenc}
\usepackage[T1]{fontenc}
\usepackage{amsmath}
\usepackage{amsfonts}
\usepackage{amssymb}
\usepackage[english]{babel}
\usepackage{lmodern}
\title{Vacuum fluctuation effects due to an Abelian gauge field 
in $2+1$ dimensions, in the presence of moving mirrors}
\author{C.~D.~Fosco$^a$, M.~L.~Remaggi$^b$ and M.~C.~Rodr\'{\i}guez$^a$ 
\\
\\
\\
{\normalsize $^a\!$\it Centro At\'omico Bariloche and Instituto Balseiro}\\
{\normalsize $\!$\it Comisi\'on Nacional de Energ\'\i a At\'omica}\\
{\normalsize $\!$\it R8402AGP San Carlos de Bariloche, Argentina}\\
\\
{\normalsize $^b\!$\it Facultad de Ciencias Exactas y Naturales}\\
{\normalsize $\!$\it Universidad Nacional de Cuyo}\\
{\normalsize $\!$\it 5500 Mendoza, Argentina.}} 

\begin{document}
\date{}
\maketitle
\begin{abstract}
\noindent We study the Dynamical Casimir Effect (DCE) due to an Abelian
	gauge field in $2+1$ dimensions, in the presence of
	semitransparent, zero-width mirrors, which may move or deform in a
	time-dependent way.  We obtain general expressions for the
	probability of motion-induced pair creation, which we render in a
	more explicit form, for some relevant states of motion.
\end{abstract}
\section{Introduction}\label{sec:intro}
The Dynamical Casimir Effect (DCE) encompasses phenomena where real
particles are created out of the quantum vacuum because of the presence of
external, time-dependent conditions. The creation of particles in a
one-dimensional cavity containing a moving perfect mirror has been studied
in a pioneering work by Moore~\cite{Moore}, and subsequently by Davies and
Fulling~\cite{FulDav}.  In recent years, the DCE has received renewed
attention,  becoming a relevant topic for different, related phenomena,
like cavity quantum electrodynamics, superconducting waveguides subjected
to time-dependent boundary conditions, quantum friction, etc., (for some
reviews, see, for example~\cite{reviews}).

In this article, we consider the DCE for a $2+1$-dimensional system which
consists of either one or two semitransparent mirrors, in a non-trivial
state of motion, coupled to a quantum Abelian gauge field.  Abelian
gauge field theories in $2+1$ dimensions play a special role in Quantum
Field Theory models which are of relevance to Condensed Matter Physics
applications~\cite{Fradkin:1991nr,K1,K2}, in continuum quantum field
theory effective descriptions.  It is our aim to consider the phenomenon of
motion induced radiation in that sort of system, because of its potential
relevance in models, besides its intrinsic, theoretical interest.
We recall that motion induced radiation with non-perfect mirrors has already
been considered, as in Ref.~\cite{Barton1}, for a mirror in
nonrelativistic motion in $1+1$ dimensions. Other models have been
considered by several authors~\cite{gensemi}.  

It is our intention to extend here the idea of~\cite{Fosco:2017jjf} to the
case of an Abelian gauge field (rather than to a scalar field), and to
more general states of motion. In particular, we aim to allow for time-dependent
deformations of the mirrors.
We recall that the approach of~\cite{Fosco:2017jjf} consisted of
considering imperfect semitransparent mirrors undergoing accelerated
motion. The approximation involved in treating the mirrors perturbatively
allowed for disentangling the purely quantum calculation (due to the field)
from the treatment of the mirrors' motion, which can be incorporated at the
end of the calculation.

This paper is organized as follows: in Sect.~\ref{sec:them}, we introduce
the kind of model that we consider in our study, and we also set up our
notation and the conventions we have adopted. Then, in Sect.~\ref{sec:pert}, 
we evaluate the effective action and its imaginary part perturbatively in
the coupling of the mirror to the field. 
We do that for the cases of one and two mirrors, and derive
general expressions for the imaginary part of the effective action.
In Sect.~\ref{sec:res}, we evaluate the general expressions derived in
the previous section for some particular kinds of motion.

In Sect.~\ref{sec:conc}, we present our conclusions.

\section{The system}\label{sec:them}
The system that we consider throughout this paper has, as its quantum
dynamical variable, an Abelian gauge field $A_{\mu}(x)$ in $2+1$
dimensions~\footnote{Indices from the middle of the Greek alphabet
($\mu,\nu,\lambda,\ldots$) are assumed to run over the values $0$, $1$ and
$2$.}.  The dynamics of this field, and its coupling to the moving
`mirrors', will be encoded into an Euclidean action ${\mathcal S}(A)$,
for which we assume the structure: 
\begin{equation}\label{eq:defsa}
{\mathcal S}(A) \;=\; {\mathcal S}_0(A) \,+\, {\mathcal S}_I(A) \;,
\end{equation}
where ${\mathcal S}_0$ denotes the free gauge-field action:
\begin{equation}\label{eq:defs0}
{\mathcal S}_0(A) \;=\; \int d^3x \,\left[\frac{1}{4}  F_{\mu\nu} F_{\mu\nu} 
\,+\,\frac{\lambda}{2} (\partial \cdot A)^2 \right] \;,
\end{equation}
which includes a gauge-fixing term (we shall use \mbox{$\lambda \equiv
1$}), while ${\mathcal S}_I$ deals with the coupling between the field and
the mirror(s).  In our conventions, the Euclidean space-time metric is
tantamount to the identity matrix $\delta_{\mu\nu}$. There will be,
therefore, no difference between a given expression and another one
obtained by raising or lowering one (or more) of its space-time indices.

Let us now construct the explicit form of ${\mathcal S}_I$, for just a
single mirror; to consider more than one mirror, we just add analogous
terms for each one of them.  The mirrors are assumed to be localized, i.e.,
to occupy a spatial curve at any given time, and therefore ${\mathcal
S}_I(A)$ is an integral over the worldsheet(s) swept by the mirror(s) during
time evolution.  Thus, the worldsheet ${\mathcal M}$ for that mirror may be
parametrized using two coordinates $\sigma^\alpha$, as follows
\footnote{Indices from the beginning of the Greek alphabet
($\alpha,\beta,\gamma, \ldots$)  run from $0$ to $1$.}:
\begin{equation}\label{eq:defpar}
	\sigma \,\equiv\, (\sigma^0, \sigma^1) \, \to \, y^\mu(\sigma)
	\;.	
\end{equation}
Note that, for indices corresponding to the two-dimensional (generally
curved) worldsheet of the mirrors,  their raising or lowering may indeed be
relevant, since there is an induced non-trivial metric (see
(\ref{eq:defmetric}) below).

Taking into account the assumption of locality, a simple gauge and
reparametrization-invariant form for the interaction term \mbox{${\mathcal
S}_I ={\mathcal S}_{\mathcal M}(A,y)$} is the following:
\begin{equation}\label{eq:defsm}
{\mathcal S}_{\mathcal M}(A,y) \;=\; \frac{1}{4 \xi}  
\int_{\mathcal M} d^2\sigma \sqrt{g(\sigma)} \,
g^{\alpha\alpha'}(\sigma) \, g^{\beta\beta'}(\sigma) \, 
{\mathcal F}_{\alpha\beta}(\sigma) 
{\mathcal F}_{\alpha'\beta'}(\sigma) \;, 
\end{equation}
where we have introduced:
\begin{equation}
{\mathcal F}_{\alpha\beta}(\sigma) \;\equiv\; \partial_\alpha
\mathcal{A}_\beta(\sigma) -\partial_\beta {\mathcal A}_\alpha(\sigma) \;,
\end{equation}
with $\mathcal{A}_\alpha(\sigma)$ denoting the projection of $A_\mu(x)$
onto the surface ${\mathcal M}$: 
\begin{equation}
	{\mathcal A}_\alpha(\sigma) \;\equiv\; A_\mu[y(\sigma)] \;
	e^{\mu}_{\alpha}(\sigma) \;,
\end{equation}
$e^{\mu}_{\alpha}(\sigma)$ being the tangent vectors $e^\mu_\alpha(\sigma) =
\partial y^\mu (\sigma)/\partial \sigma^\alpha $. Indices
corresponding to objects living on ${\mathcal M}$ are raised or lowered
with the induced metric tensor:
\begin{equation}\label{eq:defmetric}
g_{\alpha\beta}(\sigma) \;=\; e^\mu_\alpha(\sigma) e^\mu_\beta(\sigma) \;,
\end{equation}
and $g(\sigma) \equiv \det[g_{\alpha\beta}(\sigma)]$.

On the other hand, the  constant $\xi$ (which has the dimensions of a mass)
controls the strength of the boundary conditions; namely, $\xi \to 0$
corresponds to a perfect conductor, and $\xi \to \infty$ to no boundary
conditions being imposed on ${\mathcal M}$. Imperfect boundary conditions shall
mean a non-vanishing, finite value for $\xi$. 

A relationship that becomes useful in the forthcoming derivations, is that 
${\mathcal S}_{\mathcal M}$ may be shown to be equivalent to:
\begin{equation}\label{eq:siaeq}
{\mathcal S}_{\mathcal M}(A,y) \;=\; \frac{1}{2 \xi}  
\int_{\mathcal M} d^2\sigma \sqrt{g(\sigma)} \,  \big( \hat{n}_\mu(\sigma)
\tilde{F}_\mu[y(\sigma)] \big)^2 \;, 
\end{equation}
where $\tilde{F}_\mu(x) = \epsilon_{\mu\nu\lambda} \partial_\nu
A_\lambda(x)$, and  $\hat{n}_\mu(\sigma)$ is the unit normal to the
surface:
\begin{equation}
	\hat{n}_\mu(\sigma) \;=\; \frac{N_\mu(\sigma)}{\sqrt{N^2(\sigma)}}
	\;\;,\;\;\;  N_\mu(\sigma) \,=\, \frac{1}{2}\epsilon^{\alpha\beta}
	\, \epsilon_{\mu\nu\lambda} e^\nu_\alpha(\sigma) 
e^\lambda_\beta(\sigma) \;,
\end{equation}
(it is straightforward to verify that $\sqrt{N^2(\sigma)} =
\sqrt{g(\sigma)}$).

In the case of two mirrors, denoted by $L$ and $R$, rather than
${\mathcal S}_I = {\mathcal S}_{\mathcal M}$ as before, we shall have: 
\begin{equation}\label{eq:defsilr}
	{\mathcal S}_I(A) \;=\; {\mathcal S}_L(A, y_L) \,+\, {\mathcal
	S}_R(A, y_R)  \;,
\end{equation}
where we have introduced two parametrizations, denoted respectively by
$y_L^\mu(\sigma_L)$ and $y_R^\mu(\sigma_R)$ for the respective mirrors.
Besides having not necessarily equal coupling constants $\xi_L$, $\xi_R$,
the actions are assumed to have exactly the same structure as ${\mathcal
S}_{\mathcal M}(A,y)$.

The observable we shall be concerned with here is the pair-creation probability
${\mathcal P}$, which in turn may be obtained from $\Gamma$, the effective action
obtained by integrating out the vacuum fluctuations of $A$ in the presence of the mirror(s):
\begin{equation}\label{eq:defgamma}
e^{- \Gamma} \;=\; {\mathcal Z} \;=\; \int {\mathcal D}A \;
e^{-{\mathcal S}(A)} \;. 
\end{equation}
By its very definition, $\Gamma$ is a functional of the geometry of the
mirror(s), and a function of the constants that control the strength of the
coupling between them and the field.

From (\ref{eq:defgamma}), we see that the effective action may be written
as follows:
\begin{equation}
	e^{- \Gamma} \;=\; e^{- \Gamma_0} \; e^{- \Gamma_I} \;, 
\end{equation}
where
\begin{equation}\label{eq:defgamma_0}
e^{- \Gamma_0} \; \equiv \; {\mathcal Z}_0 \;=\; \int {\mathcal D}A \;
e^{-{\mathcal S}_0(A)} \;, 
\end{equation}
is the effective action in the absence of the mirror(s), and:
\begin{equation}
e^{- \Gamma_I} \; \equiv \; \langle e^{- {\mathcal S}_I} \rangle \;, 
\end{equation}
where we have introduced the $\langle \cdot \rangle$ symbol to denote 
functional averaging, with a Gaussian weight determined by the free action, namely:
\begin{equation}
\left\langle \ldots \right\rangle \;\equiv\; 
\frac{\int {\mathcal D}A \;\ldots\; e^{-{\mathcal S}_0(A)}}{\int {\mathcal D}A \; 
e^{-{\mathcal S}_0(A)}}  \;.
\end{equation}
Therefore, since only the interaction term may produce a non-vanishing
imaginary part, the probability ${\mathcal P}$ may be written as follows:
\begin{equation}
	{\cal P} \; =\; 2\; {\rm Im}[{\Gamma_I}] \;,
\end{equation} 
where $\Gamma_I$ denotes the continuation to real time of the equally
denoted functional.

Let us consider, in the next Section, the perturbative calculation of
$\Gamma_I$ and of its imaginary part, without specifying the state of
motion of the mirror(s).
\section{Perturbation theory}\label{sec:pert}
When $\Gamma_I$ is expanded in powers of ${\mathcal S}_I$,
$\Gamma_I = \Gamma^{(1)}_I + \Gamma^{(2)}_I + \ldots$, the first and
second-order terms are given by:
\begin{equation}
\Gamma_I^{(1)} = \langle {\mathcal S}_I \rangle \;,
\end{equation}
and
\begin{equation}
\Gamma_I^{(2)} = \frac{1}{2} {\langle {\mathcal S}_I 
 \rangle}^2 - \frac{1}{2} \langle {\mathcal S}_I^2 \rangle = - \frac{1}{2}
\langle ({\mathcal S}_I - \langle {\mathcal S}_I \rangle)^2 \rangle \;.
\end{equation}
For a single mirror, $\Gamma_I \to \Gamma_{\mathcal M}$ is obtained by
making the substitution:
\mbox{${\mathcal S}_I \to {\mathcal S}_{\mathcal M}$} (with ${\mathcal
S}_{\mathcal M}$ as defined in (\ref{eq:defsm})) in the expressions
above, while for two mirrors the substitution \mbox{${\mathcal S}_I \to
{\mathcal S}_L + {\mathcal S}_R$} leads to:
\begin{equation}
	\Gamma^{(1)}_I \;\equiv\; \Gamma^{(1)}_L \,+\,\Gamma^{(1)}_R \;, \;\;\;
	\Gamma^{(2)}_I \;\equiv\; \Gamma^{(2)}_L \,+\, \Gamma^{(2)}_R
	\,+\,\Gamma^{(2)}_{LR} \;,
\end{equation}
where, in a self-explaining notation:
\begin{align}
	\Gamma^{(1)}_{\stackrel{L}{R}} &\;\equiv\; \Gamma^{(1)}_{\mathcal
	M}\Big|_{{\mathcal M} \to L,R} \;, \;\; \Gamma^{(2)}_{\stackrel{L}{R}} \equiv
\Gamma^{(2)}_{\mathcal M}\Big|_{{\mathcal M} \to L,R}, \nonumber\\
\Gamma^{(2)}_{LR} &\;=\; - \langle ({\mathcal S}_L - \langle {\mathcal S}_L \rangle) 
	({\mathcal S}_R - \langle {\mathcal S}_R\rangle ) \rangle \;. 
\end{align}
In other words, to this order, we have terms that involve just one of the
mirrors, plus one which mixes both of them.
Therefore, we just need the effective action $\Gamma^{(1,2)}_{\mathcal M}$,
corresponding to an interaction term ${\mathcal S}_{\mathcal M}$, plus
$\Gamma^{(2)}_{LR}$; the remaining ones may be obtained by performing the
appropriate substitutions in a set of `independent' functionals.  

The explicit form of the independent terms we need in order to determine all
the rest (up to the second order) is:
\begin{equation}
	\Gamma_{\mathcal M}^{(1)} \;=\; \frac{1}{2 \xi}  \int_{\mathcal M} d^2\sigma
\sqrt{g(\sigma)} \,  \hat{n}_\mu(\sigma)  \hat{n}_\nu(\sigma) \langle
\tilde{F}_\mu[y(\sigma)]  \tilde{F}_\nu[y(\sigma)] \rangle \;,
\end{equation}
\begin{align}
	\Gamma_{\mathcal M}^{(2)} &= - \frac{1}{2 (2 \xi)^2}  
\int_{\mathcal M} d^2\sigma \sqrt{g(\sigma)} \,  \hat{n}_\mu(\sigma)  \hat{n}_\nu(\sigma)
\int_{\mathcal M} d^2\sigma' \sqrt{g(\sigma')} \,
\hat{n}_{\mu'}({\sigma'})  \hat{n}_{\nu'}({\sigma'}) \nonumber\\
 & \times \; \langle \; : \tilde{F}_\mu[y(\sigma)]  \tilde{F}_\nu[y(\sigma)]: \; :
\tilde{F}_{\mu'}[y (\sigma')]  \tilde{F}_{\nu'}[y (\sigma')]: \rangle \;,
\end{align}
and 
\begin{align}
	\Gamma_{LR}^{(2)} &= - \frac{1}{2\xi_L \, 2 \xi_R}  
	\int_{{\mathcal M}_L} d^2\sigma \sqrt{g_L(\sigma)} \,  \hat{n}_\mu^L(\sigma)
	\hat{n}_\nu^L(\sigma)
	\int_{{\mathcal M}_R} d^2\sigma' \sqrt{g_R(\sigma')} \,
\hat{n}^R_{\mu'}({\sigma'})  \hat{n}^R_{\nu'}({\sigma'}) \nonumber\\
 & \times \; \langle \; : \tilde{F}_\mu[y_L(\sigma)]  \tilde{F}_\nu[y_L(\sigma)]: \; :
\tilde{F}_{\mu'}[y_R (\sigma')]  \tilde{F}_{\nu'}[y_R (\sigma')]: \rangle \;,
\end{align}
where we have used the notation $:G: \,\equiv\, G - \langle G\rangle$. 

Let us now evaluate each one of the previous terms in turn.  All of them
involve the $\langle \tilde{F}_\mu(x) \tilde{F}_\nu (x')\rangle$
correlator, which may be obtained from  the gauge-field propagator. The outcome is
\begin{equation}\label{eq:defcorr}
\langle \tilde{F}_\mu(x) \tilde{F}_\nu (x') \rangle = \int \frac{d^3
	k}{(2\pi)^3} \, e^{i k \cdot (x-x')} \;\delta^\perp_{\mu\nu}(k) \;\;,
\end{equation}
where we have introduced the object:
$\delta^\perp_{\mu\nu}(k)=\delta_{\mu \nu} - \frac{k_\mu k_\nu}{k^2}$.
 
Therefore, the first-order term becomes
\begin{equation}
	\Gamma^{(1)}_I \;=\; \frac{1}{2 \xi}  
\int_{\mathcal M} d^2\sigma \sqrt{g(\sigma)} \,  \hat{n}_\mu(\sigma)  \hat{n}_\nu(\sigma)
	\int \frac{d^3 k}{(2\pi)^3} \;
\; \delta^\perp_{\mu \nu}(k) \;,
\end{equation} 
which is UV-divergent; indeed, using an Euclidean cutoff $\Lambda$, we see
that
\begin{equation}
\int_{|k|\leq \Lambda} \frac{d^3 k}{(2\pi)^3}\;
	\; \delta^\perp_{\mu \nu}(k) \; = \; \frac{\Lambda^3}{9 \pi^2}
	\,\delta_{\mu\nu} \;.
\end{equation}
Finally, 
\begin{equation}
	\Gamma_{\mathcal M}^{(1)} \;=\; \frac{\Lambda^3}{9 \pi^2 \xi} \,  
\int_{\mathcal M} d^2\sigma \sqrt{g(\sigma)} \;=\; \frac{\Lambda^3}{9 \pi^2
	\xi} \;
	{\rm area}({\mathcal M}) \;.
\end{equation} 
As indicated, it is a divergent term proportional to the area of the worldsheet. This
may be absorbed into a renormalization of the tension associated to the curve,
and therefore,  does not contribute to dissipative effects associated to the
motion of the boundary.

Let us now consider the second-order term $\Gamma_I^{(2)}$:  applying Wick's
theorem and taking into account the form of the interaction term,
\begin{align}
	\Gamma_{\mathcal M}^{(2)}  = - \frac{1}{(2 \xi)^2}  \,
	& \int_{\mathcal M} d^2\sigma  \sqrt{g(\sigma)} \,  \hat{n}_\mu(\sigma)
	\hat{n}_\nu(\sigma) \int_{\mathcal M} d^2\sigma' \sqrt{g(\sigma')}
\,  \hat{n}_{\mu'}({\sigma'})  \hat{n}_{\nu'}({\sigma'}) \nonumber\\
	& \times \, \langle \tilde{F}_\mu[y(\sigma)] \tilde{F}_{\mu'}[y (\sigma')] \rangle 
\langle \tilde{F}_\nu[y(\sigma)] \tilde{F}_{\nu'}[y (\sigma')] \rangle \;,
\end{align}
which, recalling (\ref{eq:defcorr}), can be rendered as follows: 
\begin{equation}
	\Gamma_{\mathcal M}^{(2)} \;=\;\frac{1}{2 \xi^2} \int \frac{d^3 k}{(2\pi)^3} \;
f_{\mu \nu}(-k) \; \widetilde{\Pi}_{\mu\nu ; \mu'\nu'}(k) \; f_{{\mu'} {\nu'}}(k) 
\end{equation} 
where 
\begin{equation}\label{eq:deffmunu}
f_{\mu \nu}(k) \equiv \int d^2\sigma \sqrt{g(\sigma)} \,  \hat{n}_\mu(\sigma)  \hat{n}_\nu(\sigma)
 e^{-i k \cdot y(\sigma)} \;,
\end{equation} 
and
\begin{equation}
\widetilde{\Pi}_{\mu\nu;\mu'\nu'} (k) = - \frac{1}{2} \, 
\int \frac{d^3 p}{(2\pi)^3} \; \delta^{\bot}_{\mu {\mu'}}(p)
	\delta^{\bot}_{\nu {\nu'}}(k-p) \;.
\end{equation}
An entirely analogous analysis for $\Gamma^{(2)}_{LR}$ leads to:
\begin{equation}
\Gamma_{LR}^{(2)} \;=\;\frac{1}{\xi_L \xi_R} \int \frac{d^3 k}{(2\pi)^3} \;
f^L_{\mu \nu}(-k) \; \widetilde{\Pi}_{\mu\nu,\mu'\nu'}(k) \; f^R_{\mu'\nu'}(k) 
\end{equation} 
where $f^L_{\mu\nu}$ and $f^R_{\mu\nu}$ are defined as in
(\ref{eq:deffmunu}), for the respective mirror, and {\em the same kernel\/}
$\widetilde{\Pi}_{\mu\nu,\mu'\nu'}$ as in $\Gamma_{\mathcal
M}^{(2)}$. 
Let us then consider the calculation of this kernel, which determines all the
second-order terms.  
After some algebra, we see that the structure of that object is: 
\begin{equation}
{\tilde{\Pi}}_{\mu\nu,{\mu'}{\nu'}} (k) = A\; \delta_{\mu {\mu'}}\delta_{\nu {\nu'}} 
\; + \; B_{\mu\nu,{\mu'}{\nu'}} (k)  \;,
\end{equation}
where
\begin{equation}
A \equiv - \frac{1}{6}  \int \frac{d^3 p}{(2\pi)^3} \; ,
\end{equation}
and
\begin{equation}
B_{\mu\nu,{\mu'}{\nu'}} (k) \equiv - \frac{1}{2}
\int \frac{d^3 p}{(2\pi)^3} \;
 \frac{p_{\mu} p_{\mu'}}{p^2} \frac{(k-p)_{\nu}(k-p)_{\nu'}}{(k-p)^2}\;.
\end{equation}
The term proportional to $A$ in the previous expression may also be absorbed
into a renormalization of the tension, as it was the case for the first
order calculation.  Since we are interested in dissipative effects, we
neglect this kind of contribution (which dimensional regularization
renormalizes away) from now on. 

Using dimensional regularization and introducing a Feynman parameter
representation, the $D$ dimensional version of the remaining term is:
\begin{equation}
B_{\mu\nu,{\mu'}{\nu'}} (k) = - \frac{1}{2}
\int_{0}^{1} d\alpha \int \frac{d^D p}{(2\pi)^D} \;
\frac{p_{\mu} p_{\mu'}(p+k)_{\nu}(p+k)_{\nu'}}{[(1-\alpha)p^2 + \alpha
(p+k)^2]^2}\;.
\end{equation}
For $D=3$, the result is:
\begin{equation}
\widetilde{\Pi}_{\mu\nu,{\mu'}{\nu'}}(k) \;=\; - \frac{1}{2048}
\left(\frac{2}{5}\;\delta_{\mu {\mu'}} \delta_{\nu {\nu'}}
	\;{|k|}^3 + 3\; k_{\mu} k_{\mu'} k_{\nu} k_{\nu'}\;
	\frac{1}{|k|}\right)\;,
\end{equation}
which, we recall, allows us to determine all the second-order terms, both for one or
two mirrors.

Therefore, performing the rotation back to real time, and taking imaginary
parts afterwards, we see that:
\begin{align}\label{eq:resa}
{\rm Im} \Big[\Gamma_{\mathcal M}^{(2)}\Big] &=\; \frac{1}{4096 \;
{\xi}^2} \int \frac{d^3 k}{(2\pi)^3} \; f_{\mu \nu}(-k) f_{\mu' \nu'}(k)
\nonumber\\ 
& \times \left[\frac{2}{5}\; g^{\mu {\mu'}} \, g^{\nu {\nu'}} \; {\rm Im}(|k|^3)
+ 3\;k^{\mu} k^{\mu'} k^{\nu} k^{\nu'} \, {\rm Im}( |k|^{-1} ) \right]\;,
\end{align}
while for ${\rm Im}\Big[\Gamma_{LR}^{(2)}\Big]$ we see that:
\begin{align}
{\rm Im} \Big[\Gamma_{LR}^{(2)}\Big] &=\; \frac{1}{2048 \; \xi_L \xi_R}
\int \frac{d^3 k}{(2\pi)^3} \; f^L_{\mu \nu}(-k) f^R_{\mu' \nu'}(k)
\nonumber\\ 
& \times \left[\frac{2}{5}\; g^{\mu {\mu'}} \, g^{\nu {\nu'}} \; {\rm Im}(|k|^3)
+ 3\;k^{\mu} k^{\mu'} k^{\nu} k^{\nu'} \, {\rm Im}( |k|^{-1} ) \right]\;.
\label{res1}
\end{align}

The relevant imaginary parts are:
\begin{equation}
\operatorname{Im} (|k|^3) \;=\; \theta (|k_0|-|\mathbf{k}|)\; 
(k_0^2-{|\mathbf{k}|}^2)^{3/2}\; ,
\end{equation} 
and
\begin{equation}
	\operatorname{Im} \left(|k|^{-1}\right) \;=\; \theta 
(|k_0|-|\mathbf{k}|)\;
(k_0^2-|{\mathbf{k}|}^2)^{-1/2} \; ,
\end{equation} 
\noindent where $\theta$ denotes Heaviside's step function and $\mathbf{k}\equiv 
(k_1,k_2)$.

Expressions (\ref{eq:resa}) and (\ref{res1}) may be regarded as general
results, where the motion is encoded in $f_{\mu\nu}$ and quantum effect
belong to the kernel, depending on $k$.  
\section{Examples}\label{sec:res}
Let us consider here the form adopted by the second-order contributions to
the imaginary part, in situations where one can obtain more explicit
expressions. 
\subsection{Small departures with respect to a planar world-sheet,
single mirror} 
As a first concrete example, we consider small, time-dependent departures
about a static spatial straight line. In other words, small deformations of
a planar world-sheet, which we assume to coincide with the $y^2=0$ plane:  
\begin{equation}
y^0 \;=\; x^0 \;,\;\; y^1 \;=\; x^1\;, \;\; y^2 \;=\; q(x_{\shortparallel}) 
\end{equation}
where $x_{\shortparallel} \;=\; (x^0, x^1)$, and $q(x_{\shortparallel})$
represents smalls departures from the static straight line configuration.
We thus expand $f_{\mu\nu}(k)$ in powers of $q(x_{\shortparallel})=0$,
obtaining: 
\begin{equation}
f_{\mu \nu}(k) \;=\; f_{\mu \nu}^{(0)}(k) \;+\; f_{\mu \nu}^{(1)}(k)\;+\; 
\ldots \;,
\end{equation}
where
\begin{equation}
f_{\mu \nu}^{(0)}(k) \;=\; \delta_{\mu}^2 \delta_{\nu}^2 \delta(k^0) 
\delta(k^1) \;,
\end{equation}
and
\begin{equation}
f_{\mu \nu}^{(1)}(k) \;=\; -i \; (\delta_{\mu}^2 \delta_{\nu}^2 \,  k^2 + 
\delta_{\mu}^2 \delta_{\nu}^{\alpha} \, k_{\alpha} + \delta_{\mu}^{\alpha} 
\delta_{\nu}^2 \, k^{\alpha}) \, 
\tilde{q}(k_{\shortparallel}) \;,
\end{equation}
where we have introduced the Fourier transform of the departure:
\begin{equation}
\tilde{q}(k_{\shortparallel}) \;=\; \int d^2 x_{\shortparallel} \;
q(x_{\shortparallel})\; 
e^{-i k_{\shortparallel} \cdot x_{\shortparallel}} \;.
\end{equation}
It may be verified that, up to the second order in the departure, the only
non-vanishing contribution to the imaginary part comes from using (twice)
the first-order term for $f_{\mu\nu}$ in the general expression.  We then
find the pair-creation probability ${\cal P} \; = \; 2\;
\operatorname{Im}[{\Gamma_{\mathcal M}}^{(2)}]$ to be:
\begin{align}
{\cal P} \;=\; &\frac{1}{2^{11} \; {\xi}^2} \int \frac{d^3 k}{(2\pi)^3}
\,\theta (|k_0|-|\mathbf{k}|)\
\bigg[\frac{2}{5}\; ((k^2)^2+2k_{\parallel}^2)\;
((k^0)^2-{|\mathbf{k}|}^2)^{3/2} \nonumber\\
+ \; &3 \; (k^2)^2 \; ((k^2)^2+2k_{\parallel}^2)^2\;
((k^0)^2-|{\mathbf{k}|}^2)^{-1/2}\bigg] 
|\tilde{q}(k_{\parallel})|^2 \;.
\end{align} 
Performing the $k^2$ integral, we obtain a more compact expression, depending only on
the momenta which are parallel to the space-time plane: 
\begin{equation} \label{small departures single plane}
{\cal P} \;=\; \frac{941}{2^{16} \; 5 \; {\xi}^2} 
\int \frac{d^2 k_{\parallel}}{{(2\pi)}^2 }\;\theta (|k^0| - |k^1|)\,
|k_{\parallel}|^6  \;  \big|\tilde{q}(k_{\parallel}) \big|^2 
\;,
\end{equation} 
which exhibits a power-like spectrum.

To get an even more concrete expression, we consider a case in which the
deformation of the linear boundary amounts to a standing wave.
Therefore, we chose $q(k_{\shortparallel})$ as follows:
\begin{equation}
q(k_{\shortparallel}) \;=\;  \epsilon \, \cos(\Omega x^0)\cos(p x^1)\;,
\end{equation}
where $\epsilon$, $\Omega$ and $p$ are positive constants.
Thus,
\begin{align}
\tilde{q}(k_{\shortparallel}) \;=\; 4 \pi^2 \epsilon \;
[&\delta(k^0-\Omega)\delta(k^1-p)+\delta(k^0-\Omega)\delta(k^1+p) \nonumber \\ 
+&\delta(k^0+\Omega)\delta(k^1-p)+\delta(k^0+\Omega)\delta(k^1+p)]\;.
\end{align}
Replacing the previous expression into Eq.~\ref{small departures single plane} and
integrating out $k_{\parallel}$, we see that the time and space
periodicities imply a result proportional to the total time $T$ and length
$L$ of the mirror, such that the probability per unit length and time
becomes:
\begin{equation}
\frac{\cal P}{LT} \;=\; \frac{941 \; \epsilon^2}{2^{16} \; 5 \; {\xi}^2}  \;
	\theta (\Omega -p)\,  ( \Omega^2  - p^2)^3\;.
\end{equation} 
Thus, this exhibits a threshold for the frequency of the standing wave,
related to its wave number. Since the maximum velocity $v$ of each point in the
mirror is $ v \sim \Omega \epsilon$, this threshold implies that 
$v$ should be larger (in units where the speed of light $c=1$) than the
ratio $\epsilon/\lambda$, where $\lambda$ is the wavelength. Therefore, 
 to overcome the threshold with non-relativistic speeds, the
amplitude of the wave needs to be smaller that its wavelength. Namely, 
$\epsilon p < v < 1$.

\subsection{ Standing waves with small amplitude, two mirrors} 
This example corresponds to two mirrors, and the contribution we consider
is $\Gamma^{(2)}_{LR}$. We assume for the $L$ mirror the parametrization:
\begin{equation}
y^0_L \;=\; x^0 \;,\;\; y^1_L \;=\; x^1\;, \;\; y^2_L \;=\; q_L(x_{\shortparallel}) 
\end{equation}
while for the $R$ one we include an average distance $a$:
\begin{equation}
y^0_R \;=\; x^0 \;,\;\; y^1_R \;=\; x^1\;, \;\; 
y^2_R \;=\; a \,+\,  q_R(x_{\shortparallel}) \;.
\end{equation}
To the second order (first order in each of the departures), we get:
\begin{align}
{\cal P} =\, \frac{1}{2^{11} \;{\xi_L \xi_R}} 
&\int \frac{d^3 k}{(2\pi)^3} \, \theta(|k^0|-|\mathbf{k}|)\, 
\cos(k^2 a) \; \tilde{q}_L(- k_{\parallel})
\tilde{q}_R(k_{\parallel}) \nonumber\\ 
	\times \,& \bigg[\frac{2}{5} ((k^2)^2+2 k_{\parallel}^2)
 ((k^0)^2-{|\mathbf{k}|}^2)^{3/2}  \nonumber\\
	& + 3  (k^2)^2 \; ((k^2)^2 + 2k_{\parallel}^2)^2
((k^0)^2-|{\mathbf{k}|}^2)^{-1/2}\bigg]  \;.
\end{align} 
Note the presence of the average distance $a$, inside the integrand.
We also see that, due to the presence of the Fourier transforms of both
departures, in order to have a non-vanishing result we need them to have a
non-vanishing overlap between those Fourier transforms. 
In the special case of motions which involve a
single mode this term will only be non-vanishing only if their frequency
and wave-number coincide.

For the special case of two standing waves in counterphase:
\begin{eqnarray}
	q_L(k_{\shortparallel}) &= & 4 \epsilon_L \, \cos(\Omega x^0)\cos(p x^1)
	\nonumber\\
	 q_R(k_{\shortparallel}) &= & - 4 \epsilon_R \, \cos(\Omega
	 x^0)\cos(p x^1)\;.
\end{eqnarray}
The probability in this case may be written as follows:
\begin{equation}
	\frac{\cal P}{LT} \;=\; \frac{\epsilon_L \epsilon_R}{2^{10} \, \pi \xi_L
	\xi_R}  \; \theta (\Omega -p)\,  ( \Omega^2  - p^2)^3\; 
	\varphi(\sqrt{\Omega^2 - p^2} a) \;, 
\end{equation} 
with
\begin{equation}
\varphi(x) \,=\, \int_{-1}^1 ds  \, \cos(s x) \;
\bigg[\frac{2}{5} (s^2+2) ( 1 - s^2)^{3/2} 
+ 3  s^2 \; (s^2 + 2)^2 ( 1 - s^2)^{-1/2}\bigg]  \;.
\end{equation} 
Performing the $s$ integration, we finally obtain
\begin{align}
\varphi(x) \;=\; \frac{ 3 \pi}{5 x^5}\;  \big[\,& x\, (340 - 111 x^2 + 45 x^4) J_0(x)\nonumber \\
 &+\; (-680 + 307 x^2 - 75 x^4) J_1(x)\big] \;.
\end{align} 

\subsection{Stationary waves with arbitrary amplitude} 
Let us consider here a qualitative difference that appears when one
considers stationary waves, or, more generally, a $q(x_\shortparallel)$ function
which is periodic in time and space. We assume those periods to be $\tau =
\frac{2\pi}{\Omega}$ and $\lambda = \frac{2\pi}{p}$, respectively.
This implies that $C_{\mu\nu}(k^2; x_\shortparallel)$, an object which appears
in the integrand which defines the function $f_{\mu\nu}$, is also periodic:
\begin{equation}
C^{\mu\nu}(k^2; x^0,x^1) \,\equiv\, \sqrt{g(x_\shortparallel)} \,
	\hat{n}^\mu(x_\shortparallel) \hat{n}^\nu(x_\shortparallel) e^{- i
	k^2 q(x_\shortparallel)} \, = \, C^{\mu\nu}(k^2; x^0 +\tau,x^1+\lambda) \;,
\end{equation} 
with the same periodicity as $q$.

Then $C^{\mu\nu}(k^2; x_\shortparallel)$ can be expanded in a double Fourier
series:
\begin{equation}
	C^{\mu\nu}(k^2; x_\shortparallel) \,=\, \sum_{l_\parallel} \, 
	\widetilde{C}^{\mu\nu}(k^2; l_\shortparallel) 
	\; e^{- i l_\shortparallel \cdot x_\shortparallel} \;,
\end{equation} 
where $l_\shortparallel = 2\pi (\frac{n^0}{\tau}, \frac{n^1}{\lambda})$,
with $n^0$ and $n^1$ integer numbers.
Thus,
\begin{equation}
	\widetilde{C}^{\mu\nu}(k^2; l_\shortparallel) \,=\, \frac{1}{\tau
	\lambda} \, \int_{0}^\tau dx^0 \int_0^\lambda dx^1\, 
	C^{\mu\nu}(k^2; x_\shortparallel) \; e^{i l_\shortparallel \cdot
	x_\shortparallel} \;.
\end{equation} 

Hence,
\begin{equation}
	f^{\mu\nu}(k) \;=\; (2\pi)^2 \, \sum_{n^0,n^1} \; 
	\widetilde{C}^{\mu\nu}(k^2; l_\shortparallel) \; 
	\delta(k_\shortparallel - l_\shortparallel) \;.
\end{equation} 

Even without knowing the exact form of the $\widetilde{C}^{\mu\nu}$
functions, we see that imaginary part of the effective action will be
proportional to the total time and the length of the mirror.
Besides, another qualitatively different feature has to do with the
threshold for the existence of an imaginary part. Indeed, the existence of
a series in $f^{\mu\nu}$ implies that, in order to have an imaginary part,
we need to have:
\begin{equation}
	|l^0| \; > \; |l^1| \;,
\end{equation}
or:
\begin{equation}
	\left| \frac{n^0}{n^1} \right| \; > \; \frac{\tau}{\lambda} \;=\;
	\frac{p}{\Omega} \;. 
\end{equation}
In other words, there always be non-vanishing contribution to the imaginary
part, regardless of the ratio between the wave frequency and wavelength.

Another analysis that can be done to study a configuration of standing waves result of defining two different waves, with opposite direction. Each wave has the form

\begin{equation}
q(x_{\shortparallel}) = A\, \cos(p_{\parallel} \, x_{\parallel}) ,
\end{equation}

\noindent and $p_{\parallel}=(p_0,p_1)$. In this case, if we assume that the derivative of q is small, it can be shown that the only contribution comes from $f_{22}$. In order to evaluate this function, we use the Jacobi-Anger expansion and we obtain that the imaginary part of the effective action is given by

\begin{align}
\operatorname{Im}\,[{\Gamma_I}^{(2)}] &= \frac{1}{2^{12}\, \xi^2}  \sum_{n = \infty}^{\infty} \int_{-\infty}^{\infty} \frac{d\,k^2}{2 \pi}\,\Theta [n^2\,((p^0)^2\, - \,(p^1)^2)\, -\,( k^2)^2]  \\
& \times \left[\frac{2}{5}\, J_n^2 (k^2\, A)  \, [n^2\,((p^0)^2\, - \,(p^1)^2)\, -\,( k^2)^2]^{3/2}\right.\\
&+\left.\, 3  \,\frac{(k^2)^4}{\sqrt{n^2\,((p^0)^2\, - \,(p^1)^2)\, -\, (k^2)^2}} \right] .
\end{align}

A similar phenomenon appears, of course, in the two-mirror case.

\section{Conclusions}\label{sec:conc}
We have evaluated the probability of vacuum decay, via the imaginary part
of the effective action $\Gamma$, for semitransparent mirrors in $2+1$ dimensions,
coupled to an Abelian gauge field.
We have therefore extended previous analysis to the case of non-scalar
field, and to non-rigid motions of the mirror(s). 

We obtained general expressions for the leading contribution to the
imaginary part of $\Gamma$, and more explicit ones for the case of small
amplitudes, and for standing waves. We believe that standing waves are a
natural configuration to consider, since they appear, for example, when one
deals with a string-like mirror with fixed ends.

We have shown that, when the motion is periodic both in time and space, the
imaginary part is always non-vanishing, with the threshold arising for
small amplitudes corresponding to just one of the possible processes
leading to pair creation.

\section*{Acknowledgements}
We acknowledge support by CONICET, ANPCyT and UNCuyo (Argentina).


\end{document}